\documentclass[]{pasj00}

\begin{document}
\SetRunningHead{Miyazaki et al.}{Time Lag in IDV of Sgr~A$^{\star}$}
\Received{2013/03/12}
\Accepted{2013/04/19}

\title{Time Lag in Intra-Day Variability of Sgr~A$^{\star}$ 
between the Light Curves at 90 and 102~GHz}

\author{Atsushi \textsc{Miyazaki} } 
\affil{Korean VLBI Network, Korea Astronomy and Space Science Institute, \\
776 Daedeokdae-ro, Yuseong-gu, Daejeon 305-348, Republic of Korea}\email{amiya@kasi.re.kr}

\author{Masato \textsc{Tsuboi}}
\affil{Institute of Space and Astronautical Science, 
Sagamihara, Kanagawa 252-5210, Japan\\
Department of Astronomy, The University of Tokyo, 
Bunkyo, Tokyo 113-0033, Japan}\email{tsuboi@vsop.isas.jaxa.jp}

\and

\author{Takahiro \textsc{Tsutsumi}}
\affil{National Radio Astronomy Observatory, 
1003 Lopezville Road, Socorro, NM 87801-0387, USA}\email{ttsutsum@nrao.edu}


\maketitle

\begin{abstract}
We performed the observation of the flux densities of Sgr~A$^\star$ at 90 and 102~GHz 
in order to detect the time lag between these frequencies 
using the Nobeyama Millimeter Array, which was previously reported at lower frequencies.  
We detected a radio flare during the observation period on 6 April 2005 
and calculated the \textit{z}-transformed discrete correlation function between the light curves. 
The time lag between these frequencies was not detected. 
If the expanding plasma model which explains the time lag at lower frequencies is valid, 
the light curve at 90~GHz would be delayed with respect to the one at 102~GHz. 
This result suggests that the plasma blobs ejected near the Galactic Center black hole 
may be widely diverse especially in optical thickness.  
Another possibility is that the major portion of the flux above 100~GHz 
does not originate from the blobs. 
\end{abstract}

\section{Introduction}

Sagittarius~A$^\star$ (Sgr~A$^\star$) is a compact source 
with emissions from radio to X-ray, and it is believed to be associated 
with the Galactic Center black hole (GCBH). 
Sgr~A$^\star$ is considered a laboratory for studying the activity of galactic nuclei.
In the radio regime, very long baseline interferometer (VLBI) is a powerful tool 
for exploring the structures of galactic nuclei. 
However, VLBI observations of Sgr~A$^\star$, at least up to 40~GHz, 
are not fruitful in spite of great efforts because the intrinsic structure of Sgr~A$^\star$ 
is hidden by electron scattering, as indicated by the observed size of Sgr~A$^\star$ 
obeying the inverse square law at those frequencies (e.g., \cite{shen05}).  
The VLBI technique at submillimeter wavelengths is advancing rapidly \citep{doel08}, 
but has yet to obtain an image of Sgr~A$^\star$. 
On the other hand, flux monitoring of Sgr~A$^\star$ 
at short centimeter and millimeter wavelengths 
has been useful to explore the region around GCBH. 
Intra-Day Variability (IDV), flux variability with a timescale of a few hours, 
has been observed frequently (e.g., \cite{miya04,mau05,li09}). 
The emitter of the radio flux of Sgr~A$^\star$ is in controversy 
but undoubtedly the emissions come from very near the GCBH. 
Two possibilities have been proposed as the radio emitter: 
hot plasma in the accretion disk of GCBH 
or expanding plasma ejected near GCBH. 
The Very Large Array (VLA) detected time lags of 20--30~min 
between flare peaks at 22 and 43~GHz \citep{YZ06,YZ08}.  
These time lags strongly suggest that the radio flux is emitted 
from adiabatically expanding plasma. 
Several authors have successfully applied the model of 
adiabatically expanding plasma to interpret the light curves 
of the observed IDVs \citep{YZ08,eck08,eck09,li09}. 
However, for the determination of the time lag, truly simultaneous observations 
at both frequencies were performed in only one epoch. 
For verifying such observations, additional simultaneous observations are needed, 
preferably at several different frequencies 
to provide further constraints on source structure of Sgr~A$^\star$. 
However, if the separation in observation frequencies is too wide, it may lead 
to ambiguity in identifying true counterparts among IDV at each frequency. 
Thus, we made observations at 90 and 102~GHz 
using the Nobeyama Millimeter Array (NMA), 
Nobeyama Radio Observatory (NRO)\footnote{NRO is a branch of 
the National Astronomical Observatory of Japan, National Institutes of Natural Sciences.}
to measure the time lag in radio variability between these frequencies.
NMA can observe Sgr~A$^\star$ at these two frequencies simultaneously, 
and thus should be able to robustly detect the time lag even if it is relatively small.

\section{Observation and Light Curves}

Aiming to detect a small time lag, we used NMA on 6 April 2005 
under stable clear weather to observe the flux densities of Sgr~A$^\star$ 
at 90 and 102~GHz.  NMA consists of six 10 m element antennas equipped 
with double side band (DSB) superconductor-insulator-superconductor (SIS) receivers 
with a single linear polarization feed.  
The Ultra-Wide-Band Correlator (UWBC), which has 1~GHz bandwidth, 
was employed for the backend \citep{oku00}. 
The lower and upper side band signals ($90\pm0.5$ and $102\pm0.5$~GHz)
were separated by $90^{\circ}$ phase switching to obtain simultaneous data. 
The instrumental gain and phase stability were calibrated by 
alternating observation of Sgr~A$^\star$ and NRAO530 (2.16~Jy in April 2005) 
at scan intervals of 23~min. 
The observation of Sgr~A$^\star$ was performed 
in the NMA array configuration with intermediate baselines, ``C-configuration'',  
giving a projected baseline range of $\sim4-47\:k\lambda$.  
The UV data were processed with the UVPROC-II software package 
developed at NRO \citep{tsut97}.  
To obtain the flux densities averaged every 1~min for a light curve, 
we applied a point source model to the visibility data, 
which was self-calibrated for phase, by using the MIRIAD package.  
The visibility data are restricted to the projected baseline larger 
than 25~$k\lambda$ in order to suppress the contamination 
from the extended components around Sgr~A$^\star$. 
The absolute uncertainty of the flux scaling is about 10\% in the 100~GHz band. 
In addition, the phase noise due to atmospheric fluctuations, 
which dominantly contributes to the visibility fitting error, 
and the effect of contamination from the extended components are considered.
These errors, which are estimated to be order of a few percent, 
are smaller than the flux scaling error. 
Thus the relative uncertainty of the flux density in adjacent bins 
should be lower than the absolute uncertainty. 
The signals at 90 and 102~GHz, which are simultaneously received 
by the DSB receiver, are separated 
by $90^{\circ}$ phase switching in the local oscillator as mentioned above. 
Most of signal paths at the both bands are the same, 
for example the same antenna and the same receiver system, 
and thus the gain drifts of the both bands are almost common. 
At the NMA site, the atmospheric condition was monitored 
with a 19~GHz radio seeing monitor 
using a reference signal from geostationary satellite. 
We checked the phase instability by the atmosphere with the seeing monitor 
during this observation, and the r.m.s. of 19~GHz phase was sufficiently small. 
The expected typical radio seeing in the 100~GHz band was 
less than at least 2 arcsec during the whole observing time, 
and less than 1 arcsec in the latter half of the observation.

\begin{figure}
  \begin{center}
    \FigureFile(80mm,90mm){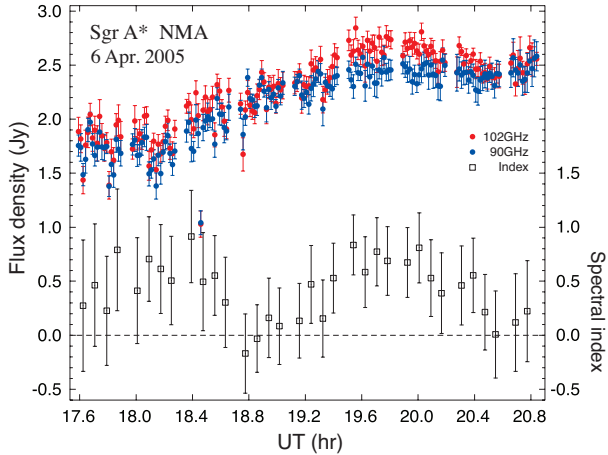}
  \end{center}
  \vspace{-3mm}
  \caption{Light curves of Sgr~A$^\star$ observed by using NMA 
  at 90 and 102~GHz on 6 April 2005. 
  Blue and red circles indicate flux densities at 90 and 102~GHz, respectively.  
  The integration time of each data point is 1~min.  
  IDV, which shows a significant increase from UT=18.2~hr to UT=19.8~hr, 
  was detected at both frequencies. 
  Open squares show the spectral index between these frequencies (averaged over 5~min).  
  }\label{fig:fig1}
\end{figure}

Figure~\ref{fig:fig1} shows the simultaneous light curves of Sgr~A$^\star$ 
at 90 and 102~GHz on 6 April 2005. 
The IDV of Sgr~A$^\star$ is clearly seen at both frequencies. 
We found that the IDV had a rising phase and broad intensity peak. 
Flux density increased from 1.7 to 2.5~Jy at 90~GHz 
and from 1.7 to 2.7~Jy at 102~GHz in 1~hr from $\sim$18.5 to $\sim$19.5~hr UT. 
The IDV timescale for a two-fold increase is estimated to be about 2~hour 
assuming that the increase has a constant gradient. 
The squares in Figure~\ref{fig:fig1} indicate the variation of spectral index 
between these frequencies in $\sim$5~min intervals. 
The spectral index is almost flat or slightly inverted, 
although the error in the first half is moderately large. 
Although the inverted spectrum of $\alpha\sim0.3\;(S_\nu\propto\nu^\alpha)$ 
has been observed in the range from several GHz to at least 100~GHz 
in the quiescent phase (e.g., \cite{tsu99,mf01}), 
the spectrum around the intensity peak at 19.5~hr 
became steep inverted up to $\alpha\sim0.8$. 
A similar inverted spectrum has also been observed in the 140~GHz band 
by using NMA \citep{miya04}.

\begin{figure}
  \begin{center}
    \FigureFile(70mm,60mm){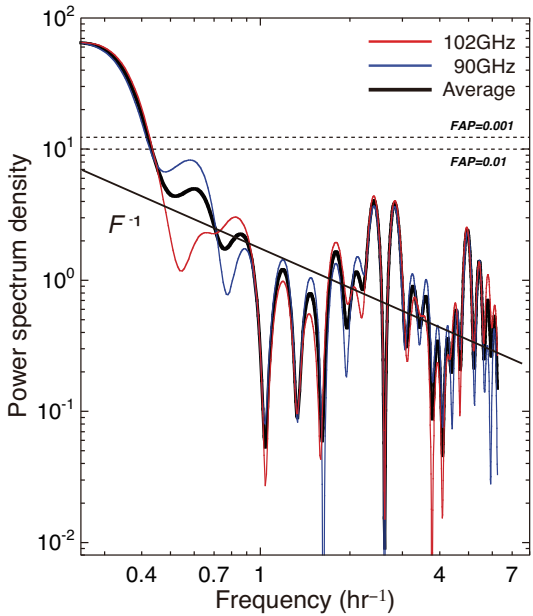}
  \end{center}
  \vspace{-3mm}
  \caption{Power spectrum density (PSD) of Sgr~A$^\star$ on 6 April 2005 
  as computed by the Lomb-Scargle periodgram method. 
  The blue and red thin curves correspond to 90 and 102~GHz, respectively. 
  The thick curve is the PSD averaged for 90 and 102~GHz.
  The straight line indicates slope of $f^{-1}$ noise. 
  Horizontal dashed lines indicate the PSDs that correspond 
  to FAPs of 0.01 and 0.001. 
  }\label{fig:fig2}
\end{figure}

We performed periodicity analyses by using the Lomb-Scargle (L-S) 
method \citep{press07} to search for any periodic behavior in the light curve. 
Figure~\ref{fig:fig2} shows the power spectrum of Sgr~A$^\star$ on 6 April 2005. 
The thick curve is the average of power-spectral densities (PSDs) at 90 and 102~GHz. 
There is an excess of PSD around the lower frequency limit. 
The excess is attributed to the rising IDV shown in Figure~\ref{fig:fig1}. 
The statistical significance of peaks in the L-S periodogram is estimated 
in terms of the false alarm probability (FAP), 
which is the probability for the PSD to exceed a certain level by chance.
FAP is given as, $FAP(> PSD)=1-(1-\exp(-PSD))^M$ 
(see eq.\ 13.8.7 in \citet{press07}), 
where $M \sim -6.362+1.193 \times N+0.00098 \times N^2$ for $N$ data points 
(see eq.\ 13 in \citet{HB86}), and $N$ is 164. 
No PSD peak with statistical significance was found 
in the periodicity of 1~hour or less. 
The PSD in the range from $f=$0.5 to 2 may show a power law behavior.
The straight line in Figure~\ref{fig:fig2} indicates the slope with power-law index 
of $f^{-1}$ for comparison. 
The slope index of the PSD is about $-1$. 
Although red noise with a slope index of $-2$ has been reported for Sgr~A$^\star$ 
in the near-infrared band (e.g., \citet{do09}), 
the slope index reported in the radio regime is $-1$ \citep{mau05}.

\section{Search for Time Lag}

\begin{figure}
  \begin{center}
    \FigureFile(75mm,90mm){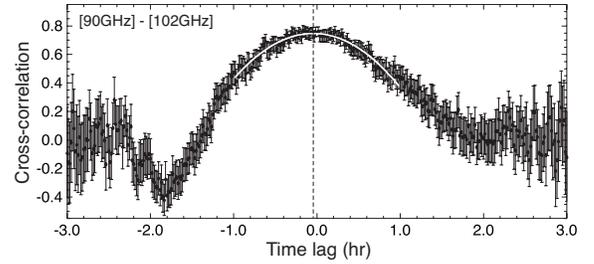}
  \end{center}
  \vspace{-3mm}
  \caption{Cross-correlation function calculated with ZDCF  
  of Sgr~A$^\star$ between 90 and 102~GHz on 6 April 2005. 
  There is a correlation function peak at time lag of approximate zero.
  The peak time lag estimated by fitting the data to a quadratic function (thick curve)
  is $-2.56\pm0.92$~min.  Vertical dashed line indicates the peak time lag. 
  }\label{fig:fig3}
\end{figure}

Cross-correlation function (CCF) analysis is a useful technique 
for finding time lags between light curves at different frequencies. 
However, since it assumes uniform sampling, 
it is not applicable to many astronomical observational data. 
The \textit{z}-transformed discrete correlation function (ZDCF) 
is a better solution to the problem of investigating correlation 
in unevenly sampled light curves \citep{alex97} and is one method 
commonly used to investigate the time correlation of active galactic nuclei.
Thus, we used ZDCF to search for a time lag 
between the light curves of Sgr~A$^\star$ at 90 and 102~GHz.
Figure~\ref{fig:fig3} shows the plots of time lag versus ZDCF between 90 and 102~GHz.  
ZDCF appears to peak at a time lag of approximate zero.
We estimated the time lag at the peak by fitting the data 
to a quadratic function in the range $\mid$time lag$\mid < 1$~hr.  
The time lag was $-2.56\pm0.92$~min. 
Each plot had a scattering within about error around a peak of ZDCF. 
Time lag with actual maximum ZDCF was offset from a peak of quadratic function shape. 
We fitted the ZDCF with a quadratic function to determine a delay against spurious peaks. 
We estimated also the centroid of ZDCF based on all points 
with correlation coefficients in excess of $0.8\;r_{max}$ 
($\tau_c = \sum_i \tau_i {\rm ZDCF}_i / \sum_i {\rm ZDCF}_i$), 
and the centroid time lag $\tau_c$ was $-2.57$~min. 
This was consistent with the value estimated by quadratic function fitting. 
A positive sign indicates the flux density at 90~GHz has a time delay
with respect to the one at 102~GHz.  The observed value indicates 
no time lag or the flux density at 102~GHz being marginally delayed 
relative to that at 90~GHz.

\begin{figure}
  \begin{center}
    \FigureFile(75mm,90mm){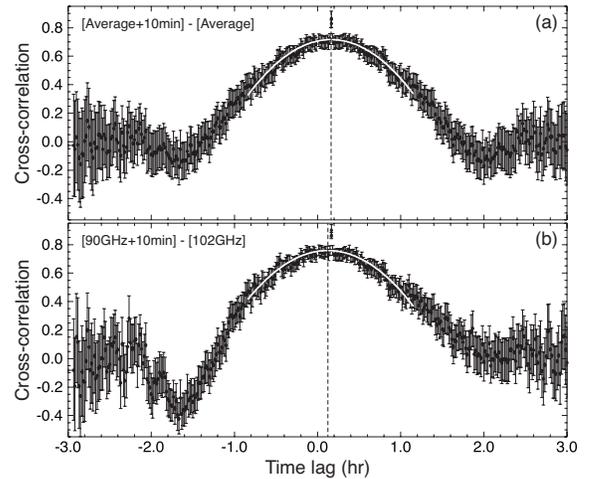}
  \end{center}
  \vspace{-3mm}
  \caption{Cross-correlation functions calculated with ZDCF 
  of the mock-up light curve data. 
  (a) Cross-correlation function between the average of the observed light curves 
  and the data that was artificially delayed 10~min. 
  The time lag, which was estimated from the peak of the best-fit quadratic 
  function (thick curve), is $9.86\pm0.88$~min. 
  (b) Cross-correlation function between the 90~GHz light curve that was artificially delayed 10~min 
  and the 102~GHz light curve. 
  The time lag estimated from the best-fit quadratic function (thick curve)
  is $7.56\pm0.85$~min. 
  Vertical dashed line in each panel indicates the peak time lag. 
  }\label{fig:fig4}
\end{figure}

We next examined whether the ZDCF algorithm would detect 
a small time lag in data with the same sampling sequence.
First, we averaged the light curves at 90 and 102~GHz to prepare mock-up data, 
and applied an artificial time delay of 10~min to prepare another set of mock-up data.
We calculated ZDCF between these two data. 
Figure~\ref{fig:fig4}a shows the cross correlation function for ZDCF. 
We estimated the time lag at the peak by fitting the mock-up data 
to a quadratic function in the range $\mid$time lag -- 10~min$\mid < 1$~hr, 
and the estimated time lag was found to be $9.86\pm0.88$~min. 
The artificial delay of 10~min was certainly detected 
with accuracy of $\sim\pm0.9$~min. 
To further verify the results, we also calculated ZDCF between a 90~GHz
light curve that was artificially delayed 10~min 
and the 102~GHz light curve (Figure~\ref{fig:fig4}b). 
By fitting to a quadratic function, the time lag at the peak ZDCF 
was estimated as $7.56\pm0.85$~min. 
As mentioned above, the 90~GHz light curve was originally found to have 
a time lag of $-2.56\pm0.92$~min with respect to the 102~GHz light curve. 
Thus, the delay for the mock-up light curves corresponds 
to $10.13\pm1.77$~min (= 7.56~min + 2.56~min). 
The time lag between these data is similar to the expected time lag.

\begin{figure}
  \begin{center}
    \FigureFile(75mm,90mm){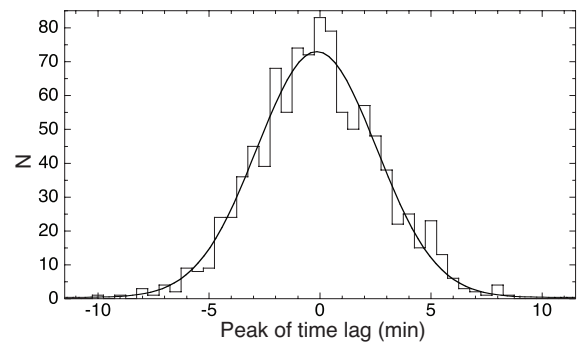}
  \end{center}
  \vspace{-3mm}
  \caption{Histogram of the ZDCF peak (time lag) 
  for mock-up light curve data with added artificial random noise.  
  The number of trials is 1000.
  The time lag follows a normal distribution with an average value of
  $-0.14\pm0.06$~min and a standard deviation of $2.69\pm0.07$~min.
  }\label{fig:fig5}
\end{figure}

In addition, we checked the accuracy of the time lag estimated 
by the ZDCF method by using Monte-Carlo simulation.
First, we prepared two mock-up data sets by the following procedure. 
An averaged light curve was constructed from the light curves at 90 and 102~GHz. 
Then, the best-fit quadratic polynomial curve was determined for the averaged light curve. 
The residuals of the light curves were calculated by subtracting the fitted curve 
from each observed light curve.  Two sets of random noise 
with the same standard deviation as the residuals were generated. 
The sets of noise were added on the fitted curve to give the mock-up data. 
Second, we calculated ZDCF between these data sets, 
and we estimated time lag by fitting to a quadratic function.
Figure~\ref{fig:fig5} shows the histogram of the time lag when this process 
was repeated 1000 times.  The histogram has an approximately normal distribution 
with an average value of $-0.14$~min and a standard deviation of 2.69~min.
Therefore, if there was a delay of a few minutes in the data, 
we would be able to detect it from the observed light curves. 
Thus, the results shown in Figure~\ref{fig:fig3} indicate that there is probably 
no time lag between the light curves of Sgr~A$^\star$ at 90 and 102~GHz.

\section{Discussion}

On the basis of VLA observations of Sgr~A$^\star$, \citet{YZ06} reported 
a time lag of $\sim$20--40~min between the flux variability at 22 and 43~GHz. 
For the origin of flare activity, two possibilities have been proposed: 
hot plasma in the accretion disk or the expanding plasma ejected near GCBH. 
The time lag that has been observed supports the expanding plasma model.
Here, following the analysis by \citet{YZ06}, 
we apply the expanding plasma model of \citet{vdL66}. 
In this model, the synchrotron-emitting plasma blob is optically thick 
when it is ejected, and becomes optically thin through adiabatic expansion. 
When the blob is optically thick, the flux density increases 
as the surface of the blob expands.
However, the flux density decreases as the optical thickness decreases 
after the transition to being optically thin. 
The frequency at which the light curve is just peaking is,
\begin{eqnarray}\label{eq:eq1}
\nu_p = \nu_0\left(\frac{R}{R_0}\right)^{1/A}, 
\end{eqnarray}
where $A=-(p+4)/(4p+6)$ and \textit{p} is the index of the relativistic electron 
energy spectrum [$\,n(E) \propto E^{-p}\,$] \citep{YZ06}. 
We assume a linear expansion model with constant expansion speed, $v_{exp}$. 
The radius of the expanding plasma is 
\begin{eqnarray}\label{eq:eq2}
R-R_0 = v_{exp} (t-t_0), 
\end{eqnarray}
where $t-t_0$ is the duration of the expansion.
Then, the relation between the peak frequency and the duration is given by 
\begin{eqnarray}\label{eq:eq3}
t-t_0 = \left[ \left(\frac{\nu_p}{\nu_0}\right)^A - 1 \right]\ \frac{R_0}{v_{exp}}. 
\end{eqnarray}
Therefore, the expected time lag between 102 and 90~GHz is calculated as 
\begin{eqnarray}\label{eq:eq4}
\Delta t_{102\mathchar`-90}\ (\mathrm{min}) =&& \left[\frac{102~\mathrm{GHz}\,^A - 90~\mathrm{GHz}\,^A}{43~\mathrm{GHz}\,^A - 22~\mathrm{GHz}\,^A} \right] \nonumber \\
&& \qquad\qquad\qquad\quad \times \Delta{T}_{43\mathchar`-22}\ (\mathrm{min}), 
\end{eqnarray}
where $\Delta{T}_{43\mathchar`-22}$ is the time lag 
between 43 and 22~GHz in minutes. 
If the time lag $\Delta{T}_{43\mathchar`-22}$ is 25~min \citep{YZ06,YZ08},
the time lag $\Delta{t}_{102\mathchar`-90}$ is expected 
to be 2.86~min at $p=2$ and 2.99~min at $p=3$.
Thus, the flux at 90~GHz is expected to be observed about 3~min behind that at 102~GHz
if the time lag between 22 and 43~GHz is the same in this observation epoch.  
However, we found no such delay in our data as shown in the previous section.
The difference between the observed time lag of $-2.56\pm0.92$~min 
and the expected time lag for $p=2$ and $\Delta{T}_{43\mathchar`-22}=25$~min 
is $\Delta{t}_{102\mathchar`-90,eff}=5.42$~min (=2.56~min + 2.86~min). 
If the electron energy spectrum is hard or steep, 
the expected time lag does not change significantly 
($\Delta{t}_{102\mathchar`-90,eff}=5.19$~min at $p=1$ 
and $\Delta{t}_{102\mathchar`-90,eff}=5.70$~min at $p=5$). 
Thus, the time lag predicted by the expanding plasma model 
is not detected in this observation epoch. 
There are at least two possible explanations of this discrepancy. 
The first possibility is that the IDV of Sgr~A$^\star$ is widely diverse, 
especially in regard to the optical thickness of the blob. 
If the blob is initially optically thin even at 100~GHz, 
the time lag between 90 and 102~GHz is not necessary.
The second possibility is that the major portion of the flux above 100~GHz 
does not originate from expanding plasma and 
instead comes from orbiting hot plasma spots on the accretion disk around GCBH. 
The variability may arise from a change in the emitting area of optically thick 
orbiting spots (e.g., \cite{BL05,li09,za10}).
Such emitter would have no time lag depending on the frequency.  
The apparent size of Sgr~A$^\star$ is increasing 
according as the well-known law: $\Delta D \propto \lambda^2$ (e.g., \cite{lo85}).  
If this phenomenon of changing the size is caused by electron scattering 
of accreting matter to the GCBH, the size indicate the diameter 
of the photosphere at the frequency.  
In the case, the lower frequency photons from the emitter pass 
for longer detour in the photosphere.  
The scenario probably explains the higher frequency photons pass 
it quickly and have a shorter time lag.  
As other model, the jet-model is proposed by \citet{fmb09}. 
On the jet-model the time lag depends on relativistic speeds of the jet, and 
the model may explain the different time lags. 
On the other hand, the fine-scale flux variations of the observed light curves  
are very complicated and can be made of consecutive flares. 
If so, even if the adiabatic expansion is at work for the individual flares, 
the light curves might show no definite time lag by blending of the neighboring flares. 
Regardless, further observations with wide frequency coverage 
are required to resolve this issue.

\section{Summary}

Using the NMA, we observed the flux densities of Sgr~A$^\star$ 
at 90 and 102~GHz on 6 April 2005 
in order to detect the time lag between these frequencies, 
and we constructed light curves covering about 3.5~hour with 1~min bins 
at both frequencies. 
We estimated ZDCF between the light curves at 90 and 102~GHz. 
The time lag derived from the peak ZDCF was $-2.56\pm0.92$~min. 
Under the expanding plasma model, 
the estimated time lag is sufficiently smaller than the time lag expected
from the previously reported observations at 22 and 43~GHz. 
In our observation data, we did not find a significant delay of the light curve at 90~GHz 
with respect to the one at 102~GHz. 
This result suggests that the plasma blobs from which the IDV originates may be diverse, 
for example, the blob may be initially optically thin even at 100~GHz. 
Another possibility is that the major portion of the flux above 100~GHz 
does not originate from expanding plasma, and 
may instead originates from hot spots on the accretion disk.

\bigskip
The authors thank the members of NMA for support during the observations.


\newpage

\end{document}